\documentclass[12pt,english]{article}

\usepackage{hyperref}

\usepackage{natbib}
\usepackage{url}
\makeatletter
\def\url@leostyle{%
 \def\UrlFont{\sf}}{\def\UrlFont{\small\ttfamily}}
\makeatother
\usepackage{graphicx}
\usepackage{babel}
\usepackage{mathrsfs}
\usepackage{amsmath}

\usepackage{authblk}

\date{}

\begin{document}

\title{Many Worlds, the Cluster-state Quantum Computer, and the
  Problem of the Preferred Basis\footnote{\textbf{Notice:} this is
  the preprint version of a work that has been accepted for
  publication in \emph{Studies in History and Philosophy of Modern
  Physics}, \href{http://dx.doi.org/10.1016/j.shpsb.2011.11.007}
  {http://dx.doi.org/10.1016/j.shpsb.2011.11.007}. Changes
  resulting from the publishing process, such as peer review,
  editing, corrections, structural formatting, and other quality
  control mechanisms are not reflected in this document. Changes
  have been made to this work since it was submitted for
  publication; in particular, the published version addresses
  one extra objection to the claims made in section
  \ref{sec:neo}.
  } \footnote{I am indebted to
  Wayne Myrvold for his comments and criticisms of earlier drafts, and
  for our many and fruitful discussions on the topic of
  this paper. I also thank Erik Curiel, Dylan Gault, and Molly Kao.}}

\author{Michael Cuffaro}
\affil{The University of Western Ontario, Department of Philosophy}

\maketitle

\thispagestyle{empty}

\section{Introduction}

The source of quantum computational `speedup'\textemdash the ability
of a quantum computer to achieve, for some problem
domains,\footnote{An important example is the factoring problem
  \citep[]{shor1997}.} a dramatic reduction in processing time over
any known classical algorithm\textemdash is still a matter of
debate. On one popular view (the `quantum parallelism
thesis'\footnote{I am indebted to \citet[]{duwell2007} for this
  label.}), the speedup is due to a quantum computer's ability to
simultaneously evaluate (using a single circuit) a function for many
different values of its input. Thus one finds, in textbooks on quantum
computation, pronouncements such as the following:

\begin{quote}
[a] qubit can exist in a superposition of states, giving a quantum
computer a hidden realm where exponential computations are possible
... This feature allows a quantum computer to do parallel computations
using a single circuit\textemdash providing a dramatic speedup in many
cases \citep[p. 197]{mcmahon2008}.
\end{quote}

\begin{quote}
Unlike classical parallelism, where multiple circuits each built to
compute $f(x)$ are executed simultaneously, here a \emph{single}
$f(x)$ circuit is employed to evaluate the function for multiple
values of $x$ simultaneously, by exploiting the ability of a quantum
computer to be in superpositions of different states
\citep[p. 31]{nielsenChuang2000}.
\end{quote}

Among textbook writers, N. David Mermin is, perhaps, the most cautious
with respect to the significance of this `quantum parallelism':

\begin{quote}
One cannot say that the result of the calculation \emph{is} 2$^n$
evaluations of $f$, though some practitioners of quantum computation
are rather careless about making such a claim. All one can say is that
those evaluations characterize the \emph{form} of the state that
describes the output of the computation. One knows what the state
\emph{is} only if one already knows the numerical values of all those
2$^n$ evaluations of $f$. Before drawing extravagant practical, or
even only metaphysical, conclusions from quantum parallelism, it is
essential to remember that when you have a collection of Qbits in a
definite but unknown state, \emph{there is no way to find out what
  that state is} \citeyearpar[p. 38]{mermin2007}.
\end{quote}

Mermin's reservations notwithstanding, the quantum parallelism thesis
is frequently associated with (and held to provide evidence for) the
many worlds explanation of quantum computation, which draws its
inspiration from the Everettian interpretation of quantum
mechanics. According to the many worlds explanation of quantum
computing, when a quantum computer effects a transition such as:
\begin{eqnarray}
\label{eqn:parallel}
\sum_{x=0}^{2^n-1} |
x \rangle | 0 \rangle \rightarrow \sum_{x=0}^{2^n-1} | x \rangle |
f(x) \rangle,
\end{eqnarray}
it literally performs, simultaneously and in different physical
worlds, local function evaluations on all of the possible values of
$x$.

The many worlds explanation of quantum computing is a very attractive
explanation of quantum speedup if one accepts the quantum parallelism
thesis, for, since the many worlds explanation of quantum computing
directly answers the question of \emph{where} this parallel processing
occurs (i.e., in distinct physical universes) in a way in which other
explanations do not, it is, arguably, the most intuitive explanation
of quantum speedup. Indeed, for some, the many worlds explanation of
quantum computing is the only possible explanation of quantum
speedup. David Deutsch, for instance, writes: ``no single-universe
theory can explain even the Einstein-Podolsky-Rosen experiment, let
alone, say, quantum computation. That is because \emph{any} process
(hidden variables, or whatever) that accounts for such phenomena
... contains many autonomous streams of information, each of which
describes something resembling the universe as described by classical
physics'' \citeyearpar[p. 542]{deutsch2010}. Deutsch issues a
challenge to those who would explain quantum speedup without many
worlds: ``[t]o those who still cling to a single-universe world-view,
I issue this challenge: Explain how Shor's algorithm works''
\citeyearpar[p. 217]{deutsch1997}.

Recently, the development of an alternative model of quantum
computation\textemdash the cluster state model\textemdash has cast
some doubt on these claims. The standard network model (which I will
also refer to as the `circuit' model) and the cluster state model are
computationally equivalent in the sense that one can be used to
efficiently simulate the other; however, while an explanation of the
network model in terms of many worlds seems intuitive and plausible,
it has been pointed out by \citet[pp. 474-475]{steane2003}, among
others, that it is by no means natural to describe cluster state
computation in this way.

While Steane is correct, I will argue that the problem that the
cluster state model presents to the many worlds explanation of quantum
computation runs deeper than this. I will argue that the many worlds
explanation of quantum computing is not only unnatural as an
explanation of cluster state quantum computing, but that it is, in
fact, incompatible with it.\footnote{My use of the word `incompatible'
  might strike some readers as a touch strong. I do not mean to convey
  by this any in-principle impossibility, however. Rather, I take it
  that any worthwhile explanation of a process should provide some
  useful insight into its workings, and should be motivated by the
  characteristics of the process, not by predilections for a
  particular type of explanation on the part of the explainer. My
  claim here is that, as I will show below, a many worlds explanation
  of cluster state quantum computing is completely unmotivated and
  useless even as a heuristic device for describing cluster state
  quantum computation, and is in this sense incompatible with it.} I
will show how this incompatibility is brought to light through a
consideration of the familiar preferred basis problem, for a preferred
basis with which to distinguish the worlds inhabited by the cluster
state neither emerges naturally as the result of a dynamical process,
nor can be chosen a priori in any principled way. In the process I
will provide a much needed exposition of cluster state computation to
the philosophical community.\footnote{Apart from very high-level
  discussions such as those found in \citet[]{steane2003}, no one, to
  my knowledge, has yet presented, to the philosophical community, a
  detailed exposition of cluster state quantum computing.}

In addition, I will argue that the many worlds explanation of quantum
computing is inadequate as an explanation of even the standard network
model of quantum computation. This is because, first, unlike its close
cousin, the neo-Everettian many worlds interpretation of quantum
\emph{mechanics},\footnote{One should be wary not to treat the
  `Everettian' interpretation of quantum mechanics as if it were a
  unified view. Rather, `Everettian' more properly describes a
  family of views (see \citealt[]{barrett2011} for a list and
  discussion of these), which includes but is not limited to Hugh
  Everett's original formulation \citep[]{everett1957}, `many minds'
  variants \citep[]{albert1988}, and `many worlds'
  variants. Belonging to the last named class are DeWitt's
  \citeyearpar{dewitt1971} original formulation, as well as the, now
  mainstream, `neo-Everettian' interpretation with which we will be
  mostly concerned in this paper. I follow Hewitt-Horsman (who
  attributes the name to Harvey Brown) in calling `neo-Everettian'
  the amalgam of ideas of
  \citet[]{zurek2003,saunders1995,butterfield2002,vaidman2008}, and
  especially \citet[]{wallace2002,wallace2003,wallace2010}.} where the
\emph{decoherence} criterion is able to fulfil the role assigned to
it, of determining the preferred basis for world decomposition with
respect to macro experience,\footnote{I should not be interpreted here
  as giving an argument \emph{for} the neo-Everettian interpretation
  of quantum mechanics. My views on the correct interpretation of
  quantum mechanics are irrelevant to this discussion. My claim is
  only that the decoherence basis is prima facie well-suited for the
  role it plays in the neo-Everettian interpretation.} the
corresponding criterion for world decomposition in the context of
quantum computing cannot fulfil this role except in an ad hoc
way. Second: alternative explanations of quantum computation exist
which, unlike the many worlds explanation, are compatible with both
the network and cluster state model.

The quantum parallelism thesis, and the many worlds explanation of
quantum computation that is so often associated with it, are
undoubtedly of great heuristic value for the purposes of algorithm
analysis and design, at least with regard to the network model. This
is a fact which I should not be misunderstood as disputing. What I am
disputing is that we should therefore be committed to the claim that
these computational worlds are, in fact, ontologically real, or that
they are indispensable for any explanation of quantum speedup.

My essay will proceed as follows. In order to exhibit the motivations
and intuitions for adopting a many worlds view of quantum computation,
I will begin, in section \ref{sec:alg}, with an example of a simple
quantum algorithm. In section \ref{sec:neo}, I will argue that,
despite its intuitive appeal, the many worlds view of quantum
computation is not licensed by, and in fact is conceptually inferior
to, the neo-Everettian version of the many worlds interpretation of
quantum mechanics from which it receives its inspiration. In section
\ref{sec:clu}, I will describe the cluster state model of quantum
computation and show how the cluster state model and the many worlds
explanation are incompatible. In section \ref{sec:net} I will argue,
based on the conclusions of sections \ref{sec:neo} and \ref{sec:clu},
that we should reject the many worlds explanation of quantum
computation \emph{tout court}.

\section{A Simple Quantum Algorithm}
\label{sec:alg}

The motivation for the view that quantum computation is parallel
processing (i.e., the quantum parallelism thesis), is evident when one
considers the specification of certain existing quantum
algorithms. Consider, for instance, the following simple algorithm for
solving Deutsch's problem: the problem to determine whether a boolean
function taking one bit as input and producing one bit as output is
either constant or balanced.

Such a function is constant if it produces the same output value for
each of its inputs. If we consider the functions $f:\{0,1\}
\rightarrow \{0,1\},$ the only possible constant functions are $f(x) =
0$ and $f(x) = 1$. A balanced function, on the other hand, is one for
which the output of one half of the inputs is the opposite of the
output of the other half. For the functions $f:\{0,1\} \rightarrow
\{0,1\}$ the only possible balanced functions are the identity and
bit-flip functions, which are, respectively:

\begin{tabular}{p{6.5cm} p{6.5cm}}
$$
\begin{array}{l}
f(x) = \left\{ 
\begin{array}{ll}
0 & \mbox{if } x = 0 \\
1 & \mbox{if } x = 1 \\
\end{array} \right. \\
\end{array}
$$
& 
$$
\begin{array}{l}
f(x) = \left\{ 
\begin{array}{ll}
1 & \mbox{if } x = 0 \\
0 & \mbox{if } x = 1 \\
\end{array} \right. \\
\end{array}
$$
\\
\end{tabular}

Now classically, the only way to determine whether such a function is
balanced or constant is to test the function for each possible value
of its input (i.e., for 0 and 1). In this case, that amounts to two
function invocations. In a quantum computer, however, we can learn
whether the function is balanced or constant by evaluating the
function only \emph{once}.

To implement the quantum algorithm using the circuit
model,\footnote{The exposition which follows is similar to Mermin's
  \citeyearpar[p. 44]{mermin2007}. This is an improved version of the
  algorithm originally developed by Deutsch
  \citeyearpar[pp. 111-112]{deutsch1985}. Deutsch's original
  algorithm, unlike the one given here, works probabilistically.} we
begin by preparing two qubits\footnote{A qubit is the basic unit of
  quantum information, analogous to a classical bit. It can be
  physically realized by any two-level quantum mechanical system. Like
  a bit, it can be ``on'': $| 1 \rangle$ or ``off'': $| 0 \rangle$,
  but unlike a bit it can also be in a superposition of these values.}
(initially assigned the state $| 0\rangle$) in the following
way. First, we send them each through a NOT (i.e., a Pauli-\textbf{X})
gate, which flips the state of the qubit. We then send them each
through a Hadamard gate, which transforms the state of each qubit into
a coherent superposition of the classical states, $| 0 \rangle$ and $|
1 \rangle$:\footnote{`Logic gates' in the network model of quantum
  computation are implemented as unitary transformations. The
  Pauli-\textbf{X} transformation takes $| 0 \rangle$ to $| 1 \rangle$
  and vice versa. A Hadamard transformation takes $| 0 \rangle$ to
  $\frac{| 0 \rangle + | 1 \rangle}{\sqrt 2}$ and $| 1 \rangle$ to
  $\frac{| 0 \rangle - | 1 \rangle}{\sqrt 2}$ and vice-versa.}

\begin{eqnarray*}
(\mathbf{H} \otimes \mathbf{H})(\mathbf{X} \otimes \mathbf{X})(| 0
\rangle | 0 \rangle)
& = & (\mathbf{H} \otimes \mathbf{H})(| 1
\rangle | 1 \rangle) \\
& = & \textstyle \frac{1}{2}(| 0 \rangle - | 1 \rangle )(| 0 \rangle
- | 1 \rangle ) \\
& = & \textstyle \frac{1}{2}(| 0 \rangle | 0 \rangle - | 1 \rangle
| 0 \rangle - | 0 \rangle | 1 \rangle + | 1 \rangle | 1 \rangle )
\end{eqnarray*}

We now send the two qubits through a `black box'\textemdash a unitary
gate, $\mathbf{U}_f$, representative of the function whose character
(of being either constant or balanced) we wish to determine\textemdash
which will perform the required evaluation. We define $\mathbf{U}_f$
so that it leaves the first qubit alone but XORs the second qubit with
the result of evaluating $f$ on the value of the first qubit, i.e.:
\begin{eqnarray}
\label{eqn:unitary}
\mathbf{U}_f(| x \rangle | y
\rangle) =_{df} | x \rangle | y \oplus f(x) \rangle .
\end{eqnarray}
Thus we have:
\begin{eqnarray}
\label{eqn:blackbox}
& & \textstyle \frac{1}{2}\mathbf{U}_f(| 0 \rangle | 0 \rangle - | 1
\rangle | 0 \rangle - | 0 \rangle | 1 \rangle + | 1 \rangle | 1
\rangle ) \nonumber\\
& = & \textstyle \frac{1}{2}(| 0 \rangle | 0 \oplus f(0) \rangle - |
1 \rangle | 0 \oplus f(1) \rangle - | 0 \rangle | 1 \oplus f(0) \rangle
+ | 1 \rangle | 1 \oplus f(1) \rangle ) \\
& = & \textstyle \frac{1}{2}(| 0 \rangle | f(0) \rangle - | 1
\rangle | f(1) \rangle - | 0 \rangle | \tilde{f}(0) \rangle + | 1
\rangle | \tilde{f}(1) \rangle \nonumber,
\end{eqnarray}
where $\tilde{f}(x) =_{df} 1 \oplus f(x).$ Note how the action of the
unitary transformation gives the appearance of evaluating the function
over multiple inputs at once. Now if the function is constant, then
$f(0) = f(1)$, $\tilde{f}(0) = \tilde{f}(1),$ and the state can be
expressed as $$\textstyle\frac{1}{2}(| 0 \rangle - | 1 \rangle)(| f(0)
\rangle - | \tilde{f}(0) \rangle .$$ If the function is balanced,
$f(0) \neq f(1),$ but $f(1) = \tilde{f}(0)$ and $\tilde{f}(1) = f(0),$
thus the state can be expressed as: $$\textstyle\frac{1}{2}(| 0
\rangle + | 1 \rangle)(| f(0) \rangle - | \tilde{f}(0) \rangle )$$ If
we apply a Hadamard transformation to the first qubit then the state
goes to
\begin{eqnarray}
\label{eqn:laststep}
| 1 \rangle\textstyle\frac{1}{\sqrt 2}(| f(0) \rangle
- | \tilde{f}(0) \rangle ),\quad | 0 \rangle\textstyle\frac{1}{\sqrt
2}(| f(0) \rangle - | \tilde{f}(0) \rangle )
\end{eqnarray}
for the constant and balanced scenarios, respectively. All that is
left to do, in order to determine whether the function is constant or
balanced, is to measure the first qubit in the $\{|0\rangle ,
|1\rangle\}$ (i.e., the computational) basis.

The algorithm, taken at a glance, is:
\begin{eqnarray}
\label{eqn:algorithm}
(\mathbf{H} \otimes
\mathbf{I})\mathbf{U}_f(\mathbf{H} \otimes \mathbf{H})(\mathbf{X}
\otimes \mathbf{X})(| 0 \rangle | 0 \rangle).
\end{eqnarray}
In this way, we solve the problem with only one invocation of
$\mathbf{U}_f$. It indeed appears as though we have performed two
steps in one.

\section{Neo-Everett and Quantum Computing}
\label{sec:neo}

Algorithms like Deutsch's and more impressive algorithms like Shor's
(which appear to perform many more than two steps in one) provide
strong intuitive support for the view that quantum speedup is due to a
quantum computer's ability to simultaneously evaluate a function for
different values of its input, and from here it is not a large step to
the many worlds picture of quantum computation. It is important to
note, however, that one's conception of a world, if one elects to take
this step, cannot be the one that is licensed by the neo-Everettian
many worlds interpretation of quantum mechanics. In superpositions
such as the following, $$\frac{1}{\sqrt 2}(|\alpha\rangle\otimes
|\beta\rangle + |\gamma\rangle\otimes |\delta\rangle),$$ the
neo-Everettian interpretation will not, in general, license one to
identify each term of this superposition with a distinct world, for
such a simplistic procedure for world-identification will be
vulnerable to the so-called preferred basis objection.

The problem is usually formulated in the context of macro-worlds and
macro-objects; however we can illustrate the basic idea by means of
the following simple example related to quantum computation. The
classical value $\uparrow$ can be represented, in the computational
basis,\footnote{The computational, or classical, basis for a single
  qubit is the basis $\{| 0 \rangle, | 1 \rangle\},$ which can be used
  to represent the classical bit states $\{\uparrow,\downarrow\}$,
  where $| 0 \rangle = \left (\begin{smallmatrix} 1
  \\ 0 \end{smallmatrix}\right ),$ and $| 1 \rangle = \left
  (\begin{smallmatrix} 0 \\ 1 \end{smallmatrix}\right ).$ An
  alternative basis for computation is $\{| + \rangle, | - \rangle\},$
  where $| + \rangle = \frac{1}{\sqrt{2}}\left (\begin{smallmatrix} 1
  \\ 1 \end{smallmatrix}\right ),$ and $| - \rangle =
  \frac{1}{\sqrt{2}}\left (\begin{smallmatrix} 1
  \\ -1 \end{smallmatrix}\right).$} by a qubit in the state $| 0
\rangle$. We can also represent the same qubit from the point of view
of the $\{| + \rangle, | - \rangle\}$ basis, however,
as\footnote{Since $| + \rangle = \frac{1}{\sqrt{2}}\left
  (\begin{smallmatrix} 1 \\ 1 \end{smallmatrix}\right ) =
  \frac{1}{\sqrt 2}(| 0 \rangle + | 1 \rangle)$ and $| - \rangle =
  \frac{1}{\sqrt{2}}\left (\begin{smallmatrix} 1
  \\ -1 \end{smallmatrix}\right) = \frac{1}{\sqrt 2}(| 0 \rangle - |
  1 \rangle)$, $\frac{1}{\sqrt 2}(| + \rangle + | - \rangle) =
  \frac{1}{2}(| 0 \rangle + | 1 \rangle + | 0 \rangle - | 1 \rangle) =
  \frac{1}{2}\cdot 2 | 0 \rangle = | 0 \rangle$.}

$$\frac{1}{\sqrt{2}} (| + \rangle + | - \rangle).$$

Thus depending on the basis one selects, it will be possible to regard
the qubit as either (if we select the computational basis) in the
definite state $| 0 \rangle$, existing in one world only, or (if we
select the $\{| + \rangle, | - \rangle\}$ basis), as in a
superposition of the two states, $| + \rangle$ and $| - \rangle$, and
thus as existing in two distinct worlds. Yet there seems to be no a
priori reason why we should elect to choose one basis over the other.

Neo-Everettians (see, for instance,
\citealt[]{wallace2002,wallace2003}) attempt to eliminate the
preferred basis problem by appealing to the dynamical process of
decoherence (\emph{Cf.} \citealt[]{zurek2003}) as a way of
distinguishing different worlds from one another in the wave
function. Recall that Schr\"odinger's wave equation governs the
evolution of a closed system. In nature, however, there are no closed
systems (aside from the entire universe); all systems interact, to
some extent, with their environment. When this happens, the terms in
the superposition of states representing the system decohere and
branch off from one another. From the point of view of an observer in
a particular world, this gives the appearance of wave-function
collapse\textemdash of definiteness emerging from
indefiniteness\textemdash but unlike actual collapse (i.e., collapse
as per von Neumann's projection postulate), decoherence is an
approximate phenomenon; thus some small amount of residual
interference between worlds always remains. But from the point of view
of our experience of macroscopic objects, this is, for all practical
purposes, enough to give us the appearance of definiteness within our
own world and to distinguish, within the wave-function, macroscopic
worlds that evolve essentially independently and maintain their
identities over time. Thus, a `preferred' basis with which one can
define different worlds emerges \emph{naturally}: ``the basic idea is
that dynamical processes cause a preferred basis to emerge rather than
having to be specified a priori'' \citep[p. 90]{wallace2003}.

On the neo-Everettian view, we identify patterns which are present in
the wave-function and which are more or less stable over time in this
way with macroscopic objects such as measurement pointers, cats, and
experimenters. But note that not every such pattern is granted
ontological status; whether or not we do so depends, not just on the
process of decoherence, but also on the theoretical usefulness of
including that object in our ontology: ``the existence of a pattern as
a real thing depends on the usefulness\textemdash in particular, the
explanatory power and predictive reliability\textemdash of theories
which admit that pattern in their ontology''
\citep[p. 93]{wallace2003}. Thus, while decoherence is a necessary
condition for granting ontological status to a pattern, it is not
sufficient; we \emph{also} require that doing so is theoretically
useful and fruitful.

Returning to the quantum computer, it should be clear by now that the
neo-Everettian interpretation, as described above, cannot provide
support for the view that quantum computers simultaneously evaluate
functions for different values of their input \emph{in different
  worlds}, for as we have just seen, \emph{decoherence} determines the
basis according to which we distinguish one world from another on the
neo-Everettian interpretation. The superpositions characteristic of
quantum algorithms, however, are always \emph{coherent}
superpositions. Indeed, the maximum length of a quantum computation is
directly related to the amount of time that the system remains
coherent \citep[p. 278]{nielsenChuang2000}. According to some, in
fact, it is coherence and not parallel processing which is the real
source of quantum speedup \citep[]{fortnow2003}. Decoherence, in the
context of quantum computation, effectively amounts to noise.

It appears, then, that we require a more general criterion for
branching than decoherence if we are to accommodate quantum
computation to a many worlds picture. Thus,
\citet[]{hewittHorsman2009}, who is notable among advocates of the
many worlds explanation for presenting a positive argument for the
many worlds explanation and not a mere assertion that other
explanations are impossible, rejects the idea that decoherence is the
only possible criterion for distinguishing worlds. Worlds, for
Hewitt-Horsman, are (just as in the neo-Everettian approach), defined
as substructures within the wave-function that `for all practical
purposes' are distinguishable and stable over relevant time
scales. With regards to macro experience these relevant time scales
are long, and the point of using decoherence as an identifying
criterion for distinct worlds, according to Hewitt-Horsman, is that it
is useful for identifying stable macro-patterns over such long time
scales. But the time scales relevant to quantum computation are
generally much shorter: ``they may, indeed, be \emph{de facto}
instantaneous. However, if they are useful then we are entitled to use
them'' \citep[p. 876]{hewittHorsman2009}.

In such a situation we may, according to Hewitt-Horsman, consider
coherent superpositions as representing distinct worlds for the
purposes of characterising quantum computation. ``Defining worlds
within a coherent state in this way is a simple extension of the
FAPP$^[$\footnote{FAPP stands for `for all practical purposes'.}$^]$
principle ... If our practical purposes allow us to deal with rapidly
changing worlds-structures then we may''
\citep[p. 876]{hewittHorsman2009}. As for the preferred basis problem,
it will not arise. Just as with the neo-Everettian interpretation, in
the quantum computer we have a criterion for selecting a basis with
which to decompose the wave function; in this case the basis is that
in which the different evaluations of the function are made manifest,
i.e., the computational basis.

\begin{quote}
This fits in well with intuitions that are often expressed about the
nature of quantum computations ... There are frequently statements to
the effect that it \emph{looks like} there are multiple copies of
classical computations happening within the quantum state. If one
classical state from a decomposition of the (quantum) input state is
chosen as an input, then the computation runs in a certain way. If the
quantum input state is used then it looks as if all the classical
computations are somehow present in the quantum one. ... the
recognition of multiple worlds in a coherent states [\emph{sic.}]
seems both to be a natural notion for a quantum information theorist,
and also a reasonable notion in any situation where `relevant'
time-scales are short \citep[p. 876]{hewittHorsman2009}.
\end{quote}

Certainly it does look as if the computation is composed of many
processes executing in parallel, and plausibly it can be of some
heuristic value to think of these processes as taking place in many
worlds. With this I do not disagree. However, \emph{pace}
Hewitt-Horsman, I do not believe this is enough to justify treating
these worlds as ontologically real, for unlike the criterion of
decoherence with respect to macro experience, Hewitt-Horsman's
criterion for distinguishing worlds in the context of quantum
computation seems quite ad hoc. Declaring that the preferred basis is
the one in which the different function evaluations are made manifest
is like declaring that the preferred basis with respect to macro
experience is the one in which we can distinguish classical states
from one another. But it is, in fact, a rejection of such reasoning
that leads to decoherence as a criterion for world-identification in
the first place. The decoherence basis, on the neo-Everettian view, is
not simply picked from among many possible bases as the one which
serves to capture our experience of definiteness at the
macro-level. To do so would be to commit the same sin (by
neo-Everettian lights) that is committed by other interpretations of
quantum mechanics such as Bohmian mechanics or GRW theory. This is the
sin of adding extra elements to the formalism of quantum theory in
order to preserve classicality at the macroscopic level. For the
neo-Everettian, in contrast, decoherence is appealed to as a known
physical process that \emph{in fact} gives rise to\textemdash and even
then only approximately\textemdash the appearance of distinct
classical worlds \citep[Cf.][pp. 55, 63-65]{wallace2010}. The point of
using decoherence as a criterion for distinguishing worlds is not to
save the appearance of classicality, but rather to \emph{explain} why
we experience the world classically, in this case\footnote{Note that I
  am not taking sides here in the debate over whether it is necessary
  to appeal to causes in such explanations.} by appealing to a
physical process that gives rise to our experience. The choice of the
computational basis as the basis within which different worlds are to
be distinguished, however, fulfils no such explanatory role. It does
not serve to explain the appearance of parallel classical
computation. It only declares, based on a particular priveleged
description of the computation, that parallel computation is occurring
in many worlds.\footnote{I should mention that Wallace, who I am
  taking as representative of the neo-Everettian interpretation of
  quantum mechanics, does seem to cautiously endorse a many worlds
  explanation for \emph{some} quantum algorithms: ``There is no
  particular reason to assume that \emph{all} or even \emph{most}
  interesting quantum algorithms operate by any sort of `quantum
  parallelism' ... But Shor's algorithm, at least, does seem to
  operate in this way'' \citep[p. 70, n. 17]{wallace2010}. Wallace has
  also made similar remarks in informal correspondance. But whatever
  Wallace's views on quantum computation are, they are obviously
  separable from his views on world decomposition for
  macro-phenomena.}

An advocate of the many worlds explanation might make the following
rejoinder: the computational process, considered as a whole, is just
as empirically well-established as the decoherence process is (we know
that a computation has taken place since we have the result). And just
as the decoherence process gives rise to parallel autonomously
evolving decoherent worlds which are (approximately) diagonal in the
decoherence basis, the computational process gives rise to parallel
autonomously evolving computational worlds which are diagonal in the
computational basis. Thus the computational process gives rise to and
therefore explains the computational worlds that make up the
computation just as well as the decoherence process explains the
decoherent worlds that make up classical experience.

This response is problematic, however, for it is the computation
itself, in particular what distinguishes it from classical
computation, that we are seeking an explanation for. The many worlds
explanation of quantum computation promises to explain quantum
computation in terms of many worlds, but on this response it appears
that we need to appeal to the computation in order to explain these
many worlds in the first place. This seems circular, and even if the
case can be made that it is not, the response fails to consider that,
as the Mermin quote with which I began this paper makes clear,
appearances can be misleading: we must be very cautious when
describing the quantum state characterising a computation. In
particular, we must be cautious when inferring from the form of the
state that describes the computation to the content of that state. For
instance, as Steane \citeyearpar[p. 473]{steane2003} has pointed out,
according to the Gottesman-Knill theorem, an important class of
quantum gates\textemdash the so-called Clifford-group gates, which
include the Hadamard, Pauli, and CNOT gates\textemdash can be
simulated in polynomial time by a classical probabilistic computer
\citep[p. 464]{nielsenChuang2000}. This is interesting, since several
quantum algorithms utilise gates exclusively from this class. Thus the
appearance of quantum parallelism in these cases may be deceiving.

Even if true, the quantum parallelism thesis need not entail the
existence of autonomous local parallel computational
processes. \citet[p. 1008]{duwell2007}, for instance, illustrates this
by showing how the phase relations between the terms in a system's
wave function are crucially important for an evaluation of its
computational efficiency. Phase relations between terms in a system's
wave function, however, are global properties of the system. Thus we
cannot view the computation as consisting exclusively of local
parallel computations (within multiple worlds or not). But if we
cannot do so, then there is no sense in which quantum parallelism
uniquely supports the many worlds explanation over other
explanations.

In any case, the questionable nature of the inference from the
heuristic value of the notion of computational worlds to the
ascription of ontological reality to these worlds is one good reason
to, at the very least, be suspicious of the many worlds explanation of
quantum computing. But let us, for the sake of argument, grant the
inference. Let us focus, instead, on the antecedent clause of the
conditional; i.e., on whether it really is true that the many worlds
description of quantum computation is the most useful one
available. In the next section I will examine the recently developed
cluster state model of quantum computation. I will argue that a
description of the cluster state model in terms of many worlds is, not
only unnatural, but that such a description is incompatible with the
cluster state model. I will then argue that this undermines the
usefulness of the many worlds description of quantum computation, not
just in the cluster state model, but in general.

\section{Cluster State Quantum Computing}
\label{sec:clu}

On the cluster state model
\citep[]{raussendorf2002,raussendorf2003,nielsen2005} of quantum
computation, computation proceeds by way of a series of single qubit
measurements on a highly entangled multi-qubit state known as the
cluster state.\footnote{For this reason the model has also been given
  the name `measurement based computation'.} The cluster-state quantum
computer ($QC_\mathcal{C}$) is a universal quantum computer; it can
efficiently simulate any algorithm developed within the network
model. In fact it is computationally equivalent to the network model
in the sense that each model may be used to simulate the operation of
the other. Each qubit in the cluster has a reduced density operator of
$\frac{1}{2}\mathbf{I},$ and thus individual qubit measurement
outcomes are completely random. It is nevertheless possible to process
information on the cluster state quantum computer due to the fact that
strict correlations exist between measurement outcomes. These
correlations are progressively destroyed as the computation runs its
course.\footnote{This gives rise to a third name for this model:
  `one-way computation'.}

Since most readers are familiar with the network model, it will be
easiest to illustrate the operation of the $QC_\mathcal{C}$ by
exhibiting the way one goes about simulating a network-based algorithm
with the $QC_\mathcal{C}$. In the network model, single-qubit gates
can, in general, be thought of as rotations of the Bloch sphere. For
example, the Pauli $X$, $Y$, and $Z$ gates can be thought of as
rotations of the Bloch sphere through $\pi$ radians about the $x$,
$y$, and $z$ axes, respectively. Now, it is possible to simulate an
\emph{arbitrary} rotation of the Bloch sphere with the
$QC_\mathcal{C}$ by using a chain of 5 qubits as follows (\emph{Cf.}
\citealt[pp. 446-447]{raussendorf2002},
\citealt[p. 5]{raussendorf2003}). First, we consider the Euler
representation of an arbitrary rotation.\footnote{The Euler
  representation is a way to represent the general rotation of a body
  in three dimensions. The procedure to achieve such a general
  rotation consists of three steps: a rotation of the body about one
  of its coordinate axes, followed by a rotation about a coordinate
  axis different from the first, and then a rotation about a
  coordinate axis different from the second. We represent rotations by
  Rotation operators, and matrix multiplication is used to represent
  combinations of rotations. For example, a rotation of $\alpha$ about
  $\hat{z}$ followed by a rotation of $\beta$ about $\hat{y}$ followed
  by a rotation of $\gamma$ about $\hat{x}$ is represented by
  $R_x(\gamma)R_y(\beta)R_z(\alpha)$. The analogue of the rotation
  operator in a complex state space is the unitary operator.} This is
\begin{eqnarray}
\label{eqn:euler}
U_{Rot}[\xi,\eta,\zeta] = U_x[\zeta]U_z[\eta]U_x[\xi],
\end{eqnarray}
where the rotations about the $x$ and $z$ axes are given by

\begin{eqnarray}
U_x[\alpha] & = & \mbox{exp}\left (-i\alpha\frac{\sigma_x}{2}\right ),
\\
U_z[\alpha] & = & \mbox{exp}\left (-i\alpha\frac{\sigma_z}{2}\right ).
\end{eqnarray}

The first qubit in the chain is called the input qubit; it will
contain the state that we wish to rotate. It is thus prepared in the
state $|\psi_{in}\rangle,$ while the other four qubits in the chain
are prepared in the $|+\rangle$ state. After applying an
entanglement-generating unitary transformation to the
qubits,\footnote{The procedure for generating entanglement is
  described in \citep[pp. 3-4]{raussendorf2003}.} the first four
qubits are measured one by one in the following way. We begin by
measuring qubit 1 in basis $\mathcal{B}_1(0),$ where 0 is the
measurement angle, $\phi_j$, and the basis is calculated as
\begin{eqnarray}
\label{eqn:basiseval}
\mathcal{B}_j(\phi_j) =
\left \{\frac{|0\rangle_j +
e^{i\phi_j}|1\rangle_j}{\sqrt{2}},\frac{|0\rangle_j -
e^{i\phi_j}|1\rangle_j}{\sqrt{2}}\right \}.
\end{eqnarray}
The result of this measurement is denoted $s_1,$ where $s_j\in
\{0,1\}$ represents the result of measuring the $j^{th}$ qubit.

We now use $s_1$ to calculate the measurement basis for qubit 2, which
is $\mathcal{B}_2(-\xi(-1)^{s_1})$. Qubit 2 is then measured in this
basis and the result recorded in $s_2$, which is then used to
determine the measurement basis for qubit 3:
$\mathcal{B}_3(-\eta(-1)^{s_2}).$ We then use both $s_1$ and $s_3$ to
determine the basis to use for the measurement of qubit 4:
$\mathcal{B}_4(-\zeta(-1)^{s_1 + s_3}).$ At the end of this process,
the output of the `gate' is contained in qubit 5 (i.e., qubit 5 is in
a state that is equivalent to what would have resulted if we had
applied an actual rotation to $| \psi_{in} \rangle$), which we then
read off in the computational basis.\footnote{I have simplified this
  procedure slightly. The gate simulation actually realizes, not
  exactly $U_{Rot}$, but $U'_{Rot}[\xi,\eta,\zeta] =
  U_{\Sigma,Rot}U_{Rot}[\xi,\eta,\zeta]$, where $U_{\Sigma,Rot} =
  \sigma_{x}^{s_2 + s_4}\sigma_z^{s_1 + s_3}$ is called the random
  byproduct operator and is corrected for at the end of the
  computation \citep[p. 5]{raussendorf2003}.}

Similarly, it is possible to implement more specific 1-qubit rotations
such as the Hadamard, $\pi/2$-phase, $X$,$Y$, and $Z$ gates. 2-qubit
gates, such as the CNOT gate, can be implemented using similar
techniques \citep[pp. 4-5]{raussendorf2003} and we can combine all of
these gates together in order to simulate an arbitrary network.

To illustrate the general operation of the cluster state computer,
imagine, once again, that we are simulating a network-based quantum
algorithm. In each individual gate simulation there will be, on the
one hand, those qubits whose measurement depends on the outcomes of
one or more previous measurements for the determination of their
basis, and on the other hand, those that do not. We divide these
qubits into disjoint subsets, $Q_t$, of the cluster $\mathcal{C}$, as
follows. All qubits, regardless of which gate they belong to, which do
not require a previous measurement for the determination of their
basis are added to the class $Q_0$. We then add to $Q_1$ all qubits
which depend solely on the results of measuring qubits in $Q_0$ for
the determination of their basis. $Q_2$ comprises, in turn, all qubits
which depend on the results of measuring qubits in $Q_0 \cup Q_1$ for
the determination of their basis. And so on until we reach
$Q_{t_{max}}$.

We then begin by measuring the qubits in the set $Q_0$. We use the
outcomes of these measurements to determine the measurement bases for
the qubits to be measured in $Q_1$. Once these are measured, the
outcomes of $Q_0$ and $Q_1$ together are used to determine the
measurement bases for $Q_2$. The process continues in this fashion
until all the required qubits have been measured
\citep[p. 19]{raussendorf2003}. Note that the temporal ordering of
measurements on the cluster state will, in general, not depend on what
role\textemdash input, output, etc.\textemdash qubits have with
respect to the network model. In fact, those qubits that play the role
of gates' `output registers' will typically be among the first to be
measured \citep[p. 19]{raussendorf2003}. In general, the temporal
ordering of measurements on a $QC_\mathcal{C}$ that has been designed
to simulate a network does not mirror the temporal ordering the gates
would have had if they had been implemented as a network
\citep[p. 444]{raussendorf2002}.

At this point we must ask ourselves whether it is possible to describe
the cluster state model using a many worlds ontology. At first glance
there does not seem to be anything barring such a description in
principle. We might view each of the qubits as existing simultaneously
in multiple worlds, for example, while the computation is being
performed. But even if this were possible, it is difficult to see what
would be gained by such a description, for this is neither a natural
view of what is happening, nor a particularly useful one: in the
network model it seems natural to conceive of a unitary gate as
effecting a parallel computation by means of a transformation such as
that in equation \eqref{eqn:parallel}. But such a `step' is missing in
the cluster state model. There is nothing corresponding to such a
unitary transformation. At best we have a simulation of such a gate;
however, it is a simulation that bears no resemblance, in terms of its
physical realisation, to the corresponding network circuit. In
addition, the temporal ordering of computation in the cluster state
has little, if anything, to do with the temporal ordering present in
the simulated network. Thus there is nothing corresponding to
simultaneous function evaluation in the cluster state, for on the
cluster state model gates are only conceptual entities that one may
utilise for algorithm design. When it comes to implementation, the
logical division of the cluster into distinct gates is completely
irrelevant. Indeed, in order to characterise the cluster state model
it is not necessary to begin with the logical layout of the network
model at all, for the cluster state model is, arguably, more
effectively characterised by a graph than by a network
\citep[p. 20]{raussendorf2003}.

Far from being a natural and intuitive picture of cluster state
computation, it seems, rather, that one must work \emph{against} one's
intuition to view the cluster state model as a model of parallel
computation in many worlds, and it is hard to see how such a
description can be useful. Considerations such as these prompt Steane
to write: ``[t]he evolution of the cluster state computer is not
readily or appropriately described as a set of exponentially many
computations going on at once. It is readily described as a sequence
of measurements whose outcomes exhibit correlations generated by
entanglement'' \citeyearpar[p. 474]{steane2003}. Hewitt-Horsman, also,
reluctantly rejects the view that cluster state computation need
involve an appeal to many worlds
\citep[pp. 896-897]{hewittHorsman2009}; however, as we have seen, she
still defends the legitimacy and usefulness of describing
\emph{network} based computation in terms of many worlds and of
treating these worlds as ontologically real
\citep[pp. 890-896]{hewittHorsman2009}.

But the main problem, for one who wishes to defend a many worlds
description of the operation of the cluster state computer, is not
that such a description is neither natural nor useful. The problem is
deeper than this, for it appears that it is for all practical purposes
impossible to specify a preferred basis in which to distinguish the
worlds in which parallel computations take place in the context of the
cluster state computer. Recall that, in general, measurements in the
cluster state model are \emph{adaptive}: the basis for each
measurement will change throughout the computation and will differ
from one qubit to the next. During each time step of the computation,
the (random) results of the measurements performed in that step will
determine the measurement bases used to measure the qubits in
subsequent steps. But this random determination of measurement bases
means that there is no principled way to select a preferred basis a
priori (and even if we did, few qubits would actually be measured in
that basis), and we certainly cannot assert that there is any sense in
which a preferred basis `emerges' from this process. Thus there is no
way in which to characterise the cluster state computer as performing
its computations in many worlds, for there is no way, in the context
of the cluster state computer, to even define these worlds for the
purposes of describing the computation as a whole.

As a possible rejoinder, one might assert that the cluster state model
merely obscures the fact that the computation takes place in many
worlds, and that this would be revealed upon closer analysis by, for
instance, considering how one might go about simulating a
cluster-state computation with circuits. In fact it is possible to
simulate a cluster state using classically controlled
gates. Classically controlled gates are gates whose operation is
dependent on classical bit values (these are typically the results of
measurements). To avoid the problem of the continually changing basis,
one might take the additional step of deferring all measurements to
the end of the process. According to the principle of deferred
measurement \citep[p. 186]{nielsenChuang2000}, this is always
possible.

Such a simulation would require many more qubits and at least one more
two-qubit operation for each single qubit operation in the cluster,
however. In principle, there will be no bound to either the additional
memory or to the number of additional two-qubit gates required to
realise the simulation \citep[p. 2]{deBeaudrap2009}. Practical
methods, therefore, for simulating the cluster state with circuits
allow measurement gates to be a part of the computational process
\citep[]{childs2005,deBeaudrap2009}. They decompose the cluster state
into a series of classically controlled change of basis gates followed
by measurement gates in the standard basis. Thus this will not solve
the problem for the many worlds theorist.

But perhaps some day an ingenious theorist will find a way to simulate
cluster state computation in some other model without the use of
adaptive measurements or classically controlled change of basis
gates. What should we say then? Even in this case I think it would be
misleading to speak of the cluster state model as obscuring the fact
that many worlds are responsible for the speedup it evinces. Recall
that, for those who adhere to the many worlds explanation of quantum
computation, the motivation for describing computation as literally
happening in many worlds is that it is useful for algorithm analysis
and design to believe that these worlds are real. This motivation is
absent in the cluster state model irrespective of whether it can be
simulated in some other model. Moreover, irrespective of whether it
can be simulated in some other model, the cluster state model will, in
virtue of its unique characteristics, surely lead to new ways of
thinking about quantum computation that would not have occurred to a
theorist working only with the network model. To dogmatically hold on
to the view that, in actuality, many worlds are, at root, responsible
for the speedup evinced in the cluster state model will at best be
useless, for, as we have seen, it will not help our theorist to design
algorithms for the cluster state. At worst it will be positively
detrimental if dogmatically holding on to this view prevents our
theorist from discovering the possibilities that are inherent in the
cluster state model.

\section{The Legitimacy of the Many Worlds Explanation for the Network
Model}
\label{sec:net}

We saw, in section \ref{sec:neo}, that the many worlds explanation of
quantum computing cannot avail itself of many of the arguments in
support of the many worlds interpretation of quantum mechanics which
appeal to decoherence as a criterion for distinguishing worlds in
order to circumvent the preferred basis objection. Further, we saw
that while the decoherence basis is able to fulfil the role assigned
to it, in the many worlds interpretation of quantum mechanics, of
determining the preferred basis for world decomposition with respect
to macro experience, the corresponding criterion for world
decomposition appealed to by those who defend the many worlds
explanation of quantum computing cannot fulfil this role except in an
ad hoc way. Thus we have one reason to reject many worlds as an
explanation of the network model of quantum computation. Let us put
this consideration to one side.

We have just seen, in section \ref{sec:clu}, that the cluster state
model of quantum computation is incompatible with a many worlds
explanation of it. In spite of this, one might still wish to maintain
the view that network-based computation, at least, is computation in
many worlds. There is nothing wrong in principle with such a
stance. What makes this view problematic, however, is the fact that
the cluster-state model is computationally equivalent to the network
model. One must therefore be committed to the view that an algorithm,
when run on quantum circuits, performs its computation in many worlds;
while a simulation of the same algorithm, run on a cluster-state
computer, does not. Moreover, this is in spite of the fact that there
may be no difference in the way in which individual qubits are
physically realised in each computer.

As unfortunate as such a situation would be, it would be forced on us
if there were no other potential unifying explanations of the source
of quantum speedup available. Fortunately, however, there do exist
potential explanations for quantum speedup in the network model which,
unlike the many worlds explanation, are compatible with the cluster
state model. For instance, Steane's choice, entanglement, is one
candidate for the source of quantum speedup that has been receiving
much attention in the literature. Entanglement has been exploited, in
one way or another, in every quantum algorithm developed thus far that
has exhibited exponential speedup over classical computation, and
entanglement has been proven to be a necessary condition for quantum
speedup when using pure states \citep[]{jozsa2003}.\footnote{It has
  recently been shown, however, that one may achieve, using
  \emph{mixed} states, a modest speedup over classical computation
  even when there is no entanglement present, and there are
  indications that it is possible to achieve exponential speedup in
  this way as well. It still appears that quantum correlations of some
  kind are required for quantum speedup, which some have named
  `quantum discord'. See \citealt[]{vedral2010} for a discussion.} On
this view, quantum computers are faster than classical computers
because they perform \emph{fewer}, not more, computations. By means of
entanglement, quantum computers make it possible to manipulate the
correlations present between the logical elements of a computation
without representing these elements
themselves \cite[p. 477]{steane2003}. Jeffrey Bub's
\citeyearpar[]{bub2006,bub2010} view is somewhat similar. On this
view, again, far from computing all of the values of a function
simultaneously, quantum computers are faster because they avoid the
calculation of any values of the function whatsoever, this time by
exploiting the difference between classical and quantum logic. On
Lance Fortnow's abstract matrix formulation of the computational
complexity class associated with quantum computing, interference is
identified as the key difference between it and the corresponding
complexity class for classical probabilistic
computation.\footnote{These are: These are \textbf{BPP} (bounded-error
  probabilistic polynomial time) and \textbf{BQP} (bounded error
  quantum polynomial time). Cf. \citet{nielsenChuang2000}.} Fortnow
develops an abstract mathematical framework for representing these
computational complexity classes. In Fortnow's framework, both classes
of computation are represented by transition matrices which determine
the possible transitions between the configurations of a
nondeterministic Turing machine. This framework shows, according to
Fortnow, that the fundamental difference between quantum and classical
computation is interference: in the quantum case, matrix entries can
be negative, signifying a quantum computer's ability to realise good
computational paths with higher probability by having the bad
computational paths cancel each other out
\citep[p. 606]{fortnow2003}.

Unlike the many worlds explanation, these explanations of the source
of quantum speedup do not rely on the particular characteristics of
the network model and seem straightforwardly compatible with cluster
state computation. But the fact that the many worlds explanation of
quantum speedup is not compatible with the cluster state model, while
these other explanations of quantum speedup are, is a reason to
question its usefulness as a description of network-based quantum
computation, and thus one more reason to reject it as an explanation
of quantum speedup \emph{tout court}.

\section{Conclusion}
\label{sec:con}

I hope to have convinced the reader that, whatever the merits of the
neo-Everettian interpretation of quantum mechanics are, the many
worlds explanation of quantum computing is inadequate as a description
of either the network or the cluster state model of quantum
computation. We saw above how it depends on a suspect extension of the
methodology of the neo-Everettian approach to quantum mechanics, and
we saw how, unlike other explanations of quantum computing, it is
unable to describe the cluster state model of computation. I hope that
the reader agrees that these are convincing reasons to reject the many
worlds explanation of quantum computing.

I do not want to argue that the many worlds explanation of quantum
computation, particularly in regards to the network model, has no
heuristic value. It undoubtedly does, and thinking in this manner has
assuredly led to the development of some important quantum
algorithms. Nevertheless we should take talk of many computational
worlds with a grain of salt. Indeed, taking literally the many worlds
view of quantum computation may be positively detrimental if it
prevents us from constructing models of quantum computation, such as
the cluster state model, in the future.

\bibliographystyle{elsarticle-harv}
\bibliography{Bibliography}{}

\end{document}